\documentclass[a4paper,10pt]{article}
\usepackage[utf8]{inputenc}
\usepackage{graphicx}
\usepackage{amsmath}
\usepackage{geometry}
\usepackage{longtable}
\usepackage{array} 
\usepackage{booktabs}
\usepackage{hyperref}
\usepackage{cite}
\usepackage{caption}

\title{Evaluation Report on MCP Servers}
\author{Zhiling Luo\footnote{godot.lzl@alibaba-inc.com}, Xiaorong Shi, Xuanrui Lin, Jinyang Gao}
\date{2025-04-18}

\begin{document}

\maketitle

\section*{Executive Summary}
\begin{enumerate}
    \item There are significant differences in effectiveness and efficiency among MCP servers; using MCPs does not demonstrate a noticeable improvement compared to function call.
    \item The effectiveness of the MCP server can be enhanced by optimizing the parameters that need to be constructed by the LLM.
\end{enumerate}

\section{Introduction}
Model Context Protocol (MCP)\cite{mcp} is an open protocol that enables AI models to securely interact with local and remote resources through standardized server implementations. Thousands of MCPs have been proposed in recent months. At the same time, several model platforms, e.g. OpenAI and Alibaba-cloud announced the support of MCP in their LLM products. The outbreak of the MCP protocol has become a reality. 
To study the effectiveness and efficiency of MCP servers, we selected several widely used MCP servers and conducted an experiment to evaluate them using MCPBench on their accuracy, time, and token usage. We focused on two tasks: web search and database search. The former involves searching the internet to answer questions, while the latter entails fetching data from a database. All MCP servers were compared using the same LLM and prompt in a controlled environment. We aimed to answer the following questions:
\begin{itemize}
    \item Question 1: Are MCP servers effective and efficient in practice?
    \item Question 2: Does using MCP provide higher accuracy compared to function calls?
    \item Question 3: How to enhance the performance?
\end{itemize}
To study these questions, we propose an evaluation framework, called MCPBench, which is released at \url{https://github.com/modelscope/MCPBench}. Besides, we provide the dataset of web search and database search at the same time.

\section{Tasks and Dataset}

MCPs encompass several tasks, including computing, memory management, web searching, and database interaction. They can be categorized into two groups:

\begin{itemize}
    \item \textbf{Data Fetching}: This group retrieves data from various sources to assist LLMs in completing tasks. The effectiveness of these MCPs is determined by the accuracy of the data fetched. For example, web search MCPs utilize search engines to obtain content from websites.
    
    \item \textbf{World Changing}: These MCPs alter the state of the world. Their effectiveness hinges on whether the state has been successfully modified. For instance, GitHub MCPs can commit code that changes the data in the GitHub dataset.
\end{itemize}

Evaluating world-changing MCPs presents challenges due to the difficulty in accessing the underlying status of the data sources (e.g., GitHub dataset). Therefore, our focus will be on data fetching tasks. Specifically, the evaluation report will encompass two tasks:

\subsection*{1) Web Search}
This task takes a question as input. The LLM rewrites it into keywords or some short sentences, involving a tool that typically searches the internet and returns results to the LLM. Here is an example involving the Brave Search MCP, see Tab. \ref{tab:websearch case}.

\begin{table}
    \centering
    \caption{An example of web search task}\label{tab:websearch case}
    \begin{tabular}{|p{5cm}|p{6cm}|p{3cm}|}
        \hline
    \textbf{Input} & \textbf{Tool} & \textbf{Result}\\
    \hline
    What is the middle name of Barack Obama & brave\_web\_search \{"query":"middle name of Barack Obama"\} & Hussein\\
    \hline
    \end{tabular}
\end{table}

To eliminate biases in the dataset, we introduce multiple data sources encompassing both Chinese and English languages across various fields. 

\begin{longtable}{|p{2.5cm}|p{5cm}|p{1.5cm}|p{5cm}|}
\caption{The datasource of web search task}\label{tab:websearch source}\\
\hline
\textbf{Datasource} & \textbf{Details} & \textbf{Volume} & \textbf{Case} \\
\hline
Frames\cite{frames} & Open-source dataset & 100 & "Prompt": "As of August 1, 2024, which country hosted the FIFA World Cup the last time the UEFA Champions League was won by a club from London?", "Answer": "France" \\
\hline
News (Chinese) & Collected by cleaning data from daily Xinwen Lianbo transcripts over the past three months and processing it using reverse engineering techniques. & 100 & "Prompt": "In which city did Tesla's first energy storage super factory outside of the United States officially start production on February 11, 2025?", "Answer": "Shanghai" \\
\hline
Knowledge (Chinese) & Collected by cleaning data from knowledge-intensive websites like Wikipedia and science and technology reports, and processing it using reverse engineering techniques. & 100 & "Prompt": "What type of fish might the family Cyprinidae and the family Dace represent as their primitive types?", "Answer": "armored fish"\\
\hline
\end{longtable}

\subsection*{2) Database Search}
Database search, or database interaction, is the data retrieval task with database.
This task takes a question as input. The LLM retrieves data from the database through a database MCP server. Table \ref{tab:datasearch case} shows an example involving the MySQL MCP. We gathered datasets from various sources, see Tab \ref{tab:datasearch source}.

\begin{table}
    \centering
    \caption{An example of database search task}\label{tab:datasearch case}
    \begin{tabular}{|p{5cm}|p{6cm}|p{3cm}|}
        \hline
    \textbf{Input} & \textbf{Tool} & \textbf{Result}\\
    \hline
    Fetch the sales of Tesla Model S since 2025-01 & execute\_sql  \{"query":"Select sum(sales) from sales where series='Tesla Model S' and datetime> 2025-01"\} & 13402\\
    \hline
    \end{tabular}
\end{table}

\begin{longtable}{|p{2.5cm}|p{5cm}|p{1.5cm}|p{5cm}|}
\caption{The datasource of database search task}\label{tab:datasearch source}\\
\hline
\textbf{Datasource} & \textbf{Details} & \textbf{Volume} & \textbf{Case} \\
\hline
Car\_bi & A synthetic dataset from an automobile manufacturer datasource & 355& "Prompt": "What is the total number of orders in the South China region in February 2025?", "Answer": "0" \\\hline
SQL\_EVAL  & Sampled from SQL\_EVAL\footnote{https://github.com/defog-ai/sql-eval}. It is based off the schema from the Spider, but with a new set of hand-selected questions and queries grouped by query category. & 256 & "Prompt": Which authors have written publications in both the domain "Machine Learning" and the domain "Data Science"? "Answer": "Ashish Vaswani" \\
\hline
\end{longtable}

\section{Overview of MCP Servers}
We collected the MCP servers from GitHub\footnote{http://github.com} and Smithary.AI\footnote{https://smithery.ai}. Due to limitations of time and cost, we selected those having more call records at April 2025.

\subsection{Web Search Related MCP}
\begin{itemize}
    \item \textbf{Brave Search\cite{brave_search}:} A web and local search utilizing Brave's Search API. \\
    Source: \url{https://github.com/modelcontextprotocol/servers/tree/main/src/brave-search} \\
    Tool Name: brave\_web\_search \\
    Developer: erdnax123
    
    \item \textbf{DuckDuckGo Search Server\cite{duckduckgo_mcp_server}:} A Model Context Protocol (MCP) server providing web search capabilities through DuckDuckGo, with additional features for content fetching and parsing. \\
    Source: \url{https://github.com/nickclyde/duckduckgo-mcp-server} \\
    Tool Name: search \\
    Developer: nickclyde

    \item \textbf{Tavily MCP Server\cite{tavily_search}:} A search engine for AI agents (search + extract) powered by Tavily. \\
    Source: \url{https://github.com/tavily-ai/tavily-mcp} \\
    Tool Name: tavily-search \\
    Developer: tavily-ai

    \item \textbf{Exa Search\cite{exa_web_search}:} A search engine designed for AI by Exa. \\
    Source: \url{https://github.com/exa-labs/exa-mcp-server} \\
    Tool Name: web\_search \\
    Developer: exa-labs    

    \item \textbf{Fire Crawl Search\cite{firecrawl_search}:} Extracts web data using Firecrawl. \\
    Source: \url{https://github.com/mendableai/firecrawl-mcp-server} \\
    Tool Name: firecrawl\_search \\
    Developer: mendableai

    \item \textbf{Bing Web Search\cite{bing_web_search}:} A Model Context Protocol (MCP) server for Microsoft Bing Search API integration, enabling AI assistants to conduct web, news, and image searches. \\
    Source: \url{https://github.com/leehanchung/bing-search-mcp} \\
    Tool Name: bing\_web\_search \\
    Developer: leehanchung

    \item \textbf{BochaAI:} A search engine for AI that provides access to high-quality global knowledge from nearly ten billion web pages and ecological content sources across various fields, including weather, news, encyclopedias, healthcare, and travel. \\
    Source: Alibaba Cloud BaiLian Platform \\
    Tool Name: bocha\_web\_search \\
    Developer: Alibaba Cloud
\end{itemize}

\subsection{Web Search Related Function Calls}
For comparison, we included some function calls:
\begin{itemize}
    \item \textbf{Qwen Web Search:} Uses the SDK provided by Qwen-Max-0125\cite{qwen25} to enable online search with \texttt{extra\_body=\{"enable\_search": True\}}.
    \item \textbf{Quark Search:} A search engine that is particularly useful for searching unknown information such as weather, exchange rates, and current events. \\
    Source: Official Quark Search Plugin provided by Alibaba Cloud BaiLian Platform \\
    Tool Name: quark\_search \\
    Developer: Alibaba Cloud
\end{itemize}

\subsection{Database Search Related MCP}
\begin{itemize}
    \item \textbf{XiYan MCP Server\cite{xiyan_get_data}}: An MCP server that supports data retrieval from a database using natural language queries, powered by XiyanSQL\cite{xiyansql} as the text-to-SQL LLM. \\
    Source: \url{https://github.com/XGenerationLab/xiyan_mcp_server} \\
    Tool Name: get\_data \\
    Developer: XGenerationLab

    \item \textbf{MySQL MCP Server\cite{mysql_execute_sql}}: An implementation that facilitates secure interaction with MySQL databases. \\
    Source: \url{https://github.com/designcomputer/mysql_mcp_server} \\
    Tool Name: execute\_sql \\
    Developer: designcomputer

    \item \textbf{PostgreSQL MCP Server\cite{postgresql_query}}: A Model Context Protocol server that provides read-only access to PostgreSQL databases. \\
    Source: \url{https://github.com/modelcontextprotocol/servers/tree/main/src/postgres} \\
    Tool Name: query \\
    Developer: modelcontextprotocol
\end{itemize}

\section{Criteria for Evaluation}
\subsection{Accuracy}
Accuracy is evaluated to determine the correctness of the answer. It is assessed by an LLM-based grader. We use DeepSeek-v3\cite{deepseekai2024deepseekv3technicalreport} as the grader. The prompt is:
\begin{quote}
For the following question: \{question\} \\
Please judge whether the predicted answer is correct. It is considered correct if it addresses the key information: \\
Predicted Answer: \{prediction\} \\
Correct Answer: \{ground\_truth\} \\
Just return True or False.
\end{quote}
The accuracy of each sample is 1 if the grader replies True, and 0 otherwise. The overall accuracy is the average of all samples. Nevertheless, issues such as network disruptions, API key limitations, and program crashes prevented the complete retrieval of every sample in the dataset. Consequently, we consider another criterion in practice: accuracy of valid samples. Valid samples are those that successfully completed the entire processing pipeline and were recorded in the logs, while any incomplete or failed samples were excluded from the analysis.

\subsection{Time Consumption}
We collect the end-to-end time consumption, which includes the latency of both the LLM and the MCP server. This criteria reflects the efficiency of MCP.

\subsection{Token Consumption}
We record the pre-fill (prompt) and completion tokens used during the experiment. The token consumption will affect the incurred costs.

\subsection{Other Setups}
The experiment is executed on a server in Singapore with a dual-core CPU and 2GB RAM. The evaluation framework used is MCPBench. All MCP servers (except DuckDuckGo) are launched on the server in SSE mode. The timeout is set to 30 seconds. The system prompt is as follows:
We collect the query from \texttt{<WebSearch></WebSearch>} and send it to the MCP.
\begin{quote}
handles questions that may require web searching.
\begin{itemize}
    \item Input contains:
    \begin{itemize}
        \item The question that needs to be answered.
        \item Past search steps and their results.
    \end{itemize}
    \item Output can be either:
    \begin{itemize}
        \item If more search is needed: output in the format \texttt{<WebSearch>search\_query</WebSearch>}
        \item If the question can be answered: direct answer.
    \end{itemize}
\end{itemize}

\textbf{Example:}
Input question: "Who is the current President of the United States?" \\
- If no search has been conducted: \texttt{<WebSearch>current President of United States</WebSearch>} \\
- If sufficient information is available: "Joe Biden is the current President of the United States."
\end{quote}

For the database search, we use MySQL database whose version is 9.2, the PostgreSQL version is 15.8.

\section{Comparative Analysis}
\subsection{Question 1: Are MCP Servers Effective and Efficient in Practice?}

\begin{longtable}{p{5cm}p{2cm}p{2cm}p{2cm}p{2cm}}
\caption{The experiment results on MCPs}\label{tab:main exp}\\
\hline
\textbf{MCP Server} & \textbf{Accuracy (\%)} ↑ & \textbf{Time Consumption (s)} ↓ & \textbf{Pre-fill Token Consumption} ↓ & \textbf{Completion Token Consumption} ↓ \\
\hline
Brave Search\cite{brave_search} & 46.6 & 13.98 & 5802.35 & 236.26 \\
DuckDuckGo Search Server\cite{duckduckgo_mcp_server} & 13.62 & 64.17 & 1718.84 & 162.25 \\
Tavily MCP Server\cite{tavily_search} & 47.99 & 95.52 & 2441.73 & 196.03 \\
Exa Search\cite{exa_web_search} & 15.02 & 231.24 & 2475.24 & 190.49 \\
BoChaAI Search & 20	&35.54	&1642.71	&189.13\\
Fire Crawl Search\cite{firecrawl_search} & 58.33 & 15.44 & 1727.17 & 179.61 \\
Bing Web Search\cite{bing_web_search} & 64.33 & 12.4 & 4060.34 & 206.87 \\
\hline
\end{longtable}
The experiment results are reported in the Tab. \ref{tab:main exp}, in which we can have the following observations.
\begin{enumerate}
    \item The differences in effectiveness are significant. Based on the accuracy of valid samples, the highest accuracy is observed with Bing Web Search (64\%), while DuckDuckGo has the lowest at just 10\%, representing a difference of 54 percentage points.
    \item The differences in efficiency are even more pronounced; regarding the average time consumed for valid samples, the fastest are Bing Web Search and Brave Search, which require less than 15 seconds, while the slowest, Exa Search, takes 231 seconds (note that valid samples are cases of normal returns without timeouts, so this value is unaffected by timeouts).
    \item Token consumption is similar; based on the number of output tokens for valid samples, consumption generally falls between 150 and 250 tokens, indicating that the model consistently provides concise answers without unnecessary elaboration on its MCP usage.
\end{enumerate}

\subsection{Question 2: Does MCP Provide Higher Accuracy Compared to Function Calls?}
\begin{figure}[h]
    \centering
    \includegraphics[width=0.8\textwidth]{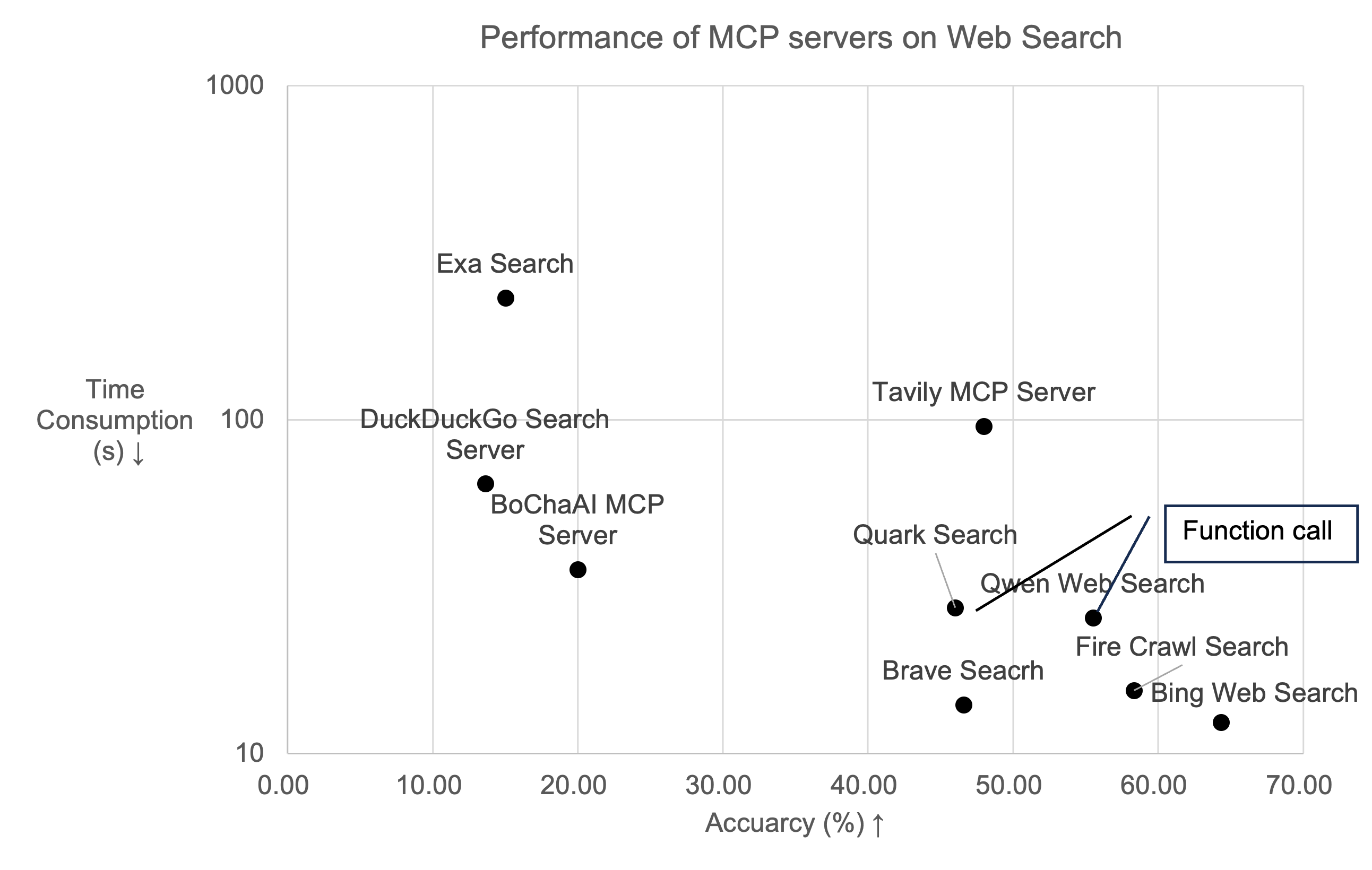} 
    \caption{The performance of MCP servers in Web search}\label{fig:main}
\end{figure}

We compared the performance of function call with the above-mentioned MCP servers. The results are shown in Figure \ref{fig:main} and Tab \ref{tab:functioncall exp}. We observe that both function calls (Qwen Web Search) and tool usage (Quark Search) exhibit competitive accuracy and time consumption. The accuracy of Qwen Web Search is 55.52\%, surpassing that of Exa Search, DuckDuckGo, Tavily, and Brave Search. There is not much difference of time consumption of function calls (Qwen Web Search and Quark Search) compared to MCP services.

\begin{longtable}{p{5.3cm}p{2cm}p{2cm}p{2cm}p{2cm}}
\caption{The experiment results of MCP and Function Call }\label{tab:functioncall exp}\\
\hline
\textbf{ } & \textbf{Accuracy (\%)} ↑ & \textbf{Time Consumption (s)} ↓ & \textbf{Pre-fill Token Consumption} ↓ & \textbf{Completion Token Consumption} ↓ \\
\hline
Quark Search (Function Call) 	& 46.00 &	27.31	&1142.21	&158.68 \\
Qwen Web Search (Function Call) 	& 55.52 &	25.48	&1149.69	&183.98 \\
Brave Search (MCP) & 46.6 & 13.98 & 5802.35 & 236.26 \\
\hline
\end{longtable}

\subsection{Question 3: How to enhance the performance?}

\begin{longtable}{p{4.5cm}p{2cm}p{2cm}p{2cm}p{2cm}}
\caption{The performance of declarative interface on MySQL MCP}\label{tab:mysql}\\
\hline
\textbf{MCP Server} & \textbf{Accuracy (\%)}↑  & \textbf{Time Consumption (s)} ↓ & \textbf{Pre-fill Token Consumption} ↓ & \textbf{Completion Token Consumption} ↓ \\
\hline
MySQL MCP Server\cite{mysql_execute_sql} & 54.73 & 4.64 & 2800 & 64.26 \\\hline
MySQL MCP Server + Declarative Interface  \\\begin{small}(XiYan MCP Server)\cite{xiyan_get_data}\end{small} & 56.06 & 6.38 & 415.52 & 44.46 \\
\hline
\end{longtable}

\begin{longtable}{p{4.5cm}p{2cm}p{2cm}p{2cm}p{2cm}}
\caption{The performance of declarative interface on PostgreSQL MCP}\label{tab:pg}\\
\hline
\textbf{MCP Server} & \textbf{Accuracy (\%)} ↑ & \textbf{Time Consumption (s)} ↓ & \textbf{Pre-fill Token Consumption} ↓ & \textbf{Completion Token Consumption} ↓ \\
\hline
PostgreSQL MCP Server\cite{postgresql_query} & 58.5&5.85	&6896.51	&103.22 \\\hline
PostgreSQL MCP Server + Declarative Interface\\\begin{small}(XiYan MCP Server)\cite{xiyan_get_data}\end{small} & 80.08&	12.87&	434.86&	97.57 \\
\hline
\end{longtable}

To address this question, we considered the database search task. The MySQL MCP server implements a straightforward encapsulation of the database connection. After configuring the database account, password, and other information, these MCPs establish a persistent connection to the database and expose the execute\_sql tool interface. When calling this tool, the model must construct a query parameter that must be an executable SQL statement. Although this simple encapsulation functions as an MCP, it assigns the most challenging part of the process—constructing the SQL query statement—to the LLM. Consequently, the success of the entire tool call heavily relies on the LLM's ability to construct SQL statements.

We introduce a Declarative Interface method to alleviate this challenge. The key idea is replace the structural parameter of MCP by a declarative interface. In other words, we use the natural language as the interface in MCP. In our experiment, we developed an updated version, called XiYan MCP server, using natural language instead of SQL. It utilizes XiYanSQL-QwenCoder-32B\footnote{https://www.modelscope.cn/models/XGenerationLab/XiYanSQL-QwenCoder-32B-2412}\cite{xiyansql} to convert natural language queries into SQL before executing comparable to the MySQL MCP server. The experiment results are shown in Table \ref{tab:mysql}. By adding a text-to-SQL model to the MCP server, it improves accuracy by 2 percentage points. 
In the PostgreSQL experiment, the optimization results is a 22-point increase \ref{tab:pg}.

\section{Case Study}
\subsection{Web Search Task}
\begin{figure}[h]
    \centering
    \includegraphics[width=1\textwidth]{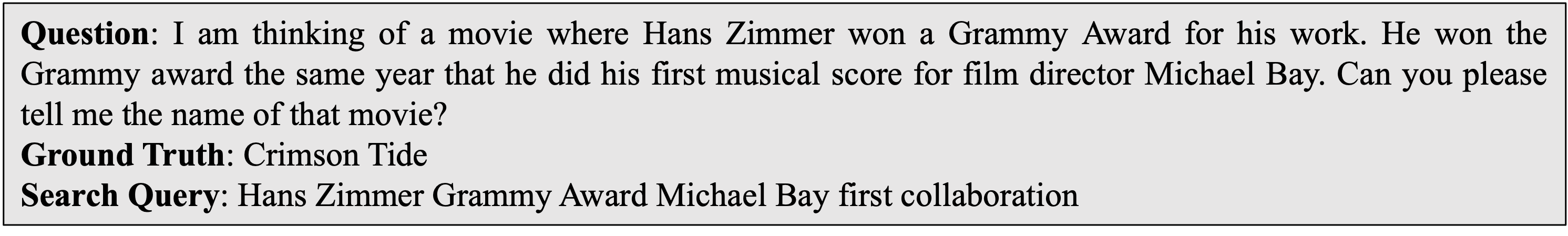} 
    \caption{The issue in the Frames dataset}\label{fig:1}
\end{figure}

For the Web Search task, we evaluate the search performance of Brave Search, BochaAI, and Qwen Web Search using the Frames dataset. The example is shown in Figure \ref{fig:1}. To ensure aesthetic consistency, the returned results were processed by removing HTML tags and truncating long outputs while maintaining the original formatting and content. Comprehensive information is available in Appendix \ref{app:web}.

\begin{figure}[h]
    \centering
    \includegraphics[width=1\textwidth]{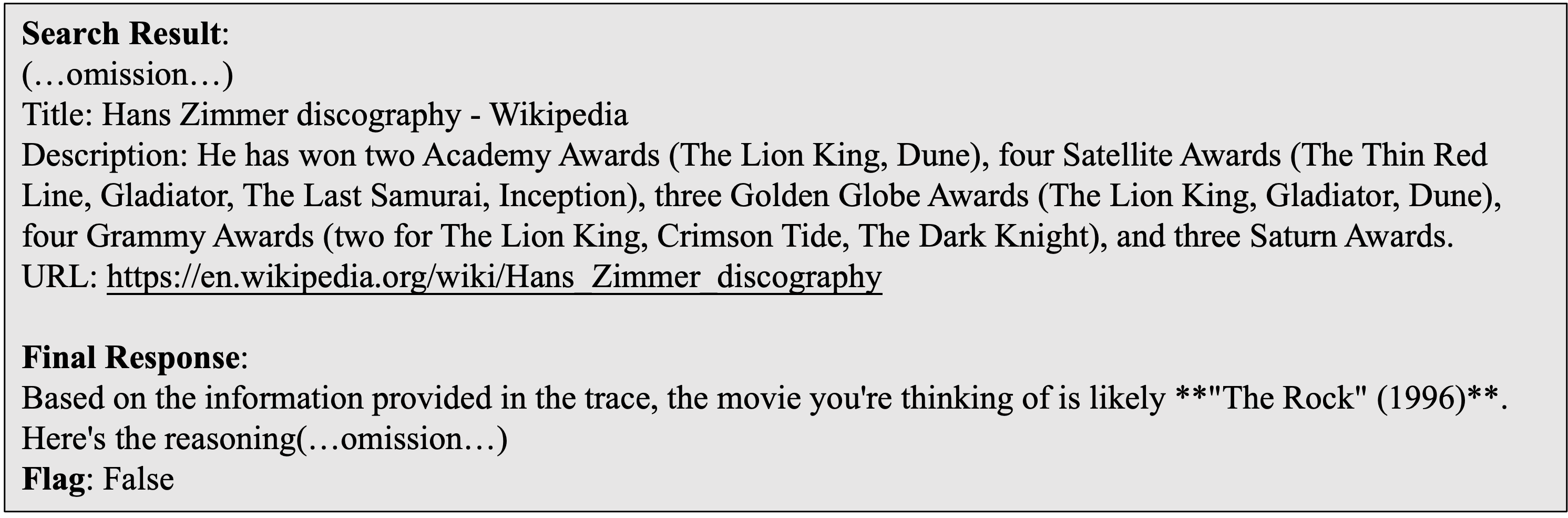} 
    \caption{The search results from Brave Search}\label{fig:2}
\end{figure}

Brave Search provides the top ten relevant Wikipedia pages, including their titles, descriptions, and URLs, as shown in Figure \ref{fig:2}. For clarity, only one of these results is displayed, which includes the ground truth (Crimson Tide). Nevertheless, the lack of detailed descriptions impedes the LLM’s ability to effectively link the question to the pertinent search result. This forces the LLM to rely solely on its internal knowledge base, ultimately leading to an incorrect response.

\begin{figure}[h]
    \centering
    \includegraphics[width=1\textwidth]{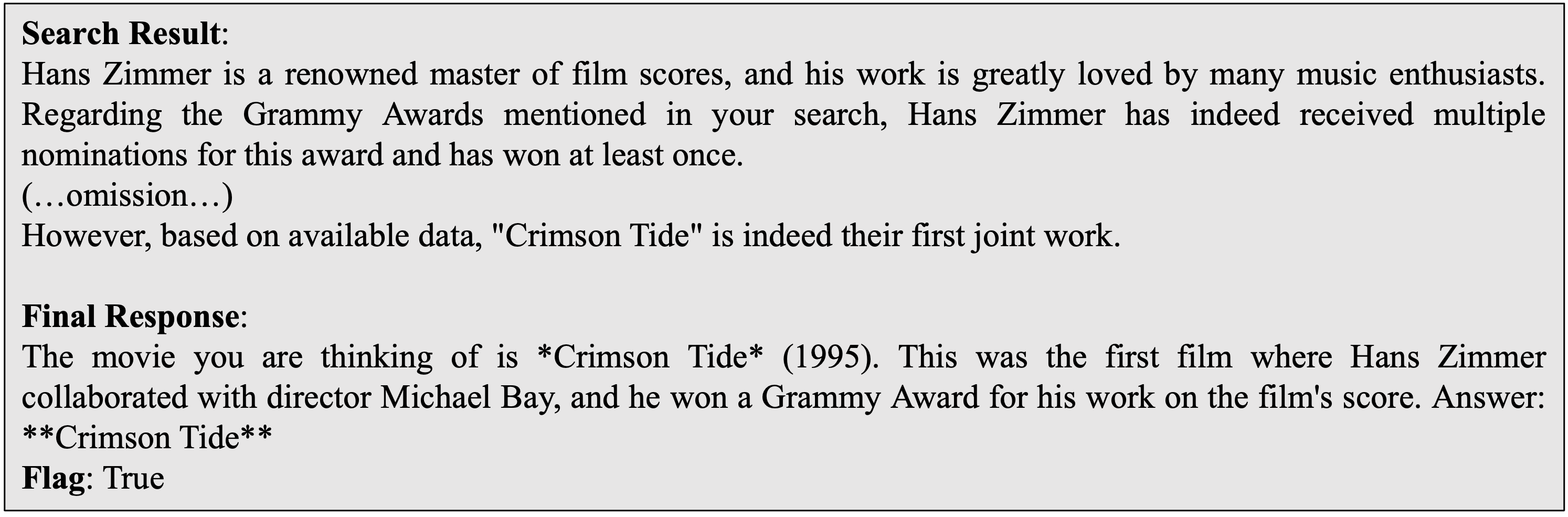} 
    \caption{The search results from BochaAI}\label{fig:3}
\end{figure}

BochaAI summarized the search results and explicitly informed the LLM that the correct answer was "Crimson Tide", as shown in Figure \ref{fig:3}. This direct approach enabled the LLM to accurately and effortlessly provide the correct answer.

\begin{figure}[h]
    \centering
    \includegraphics[width=1\textwidth]{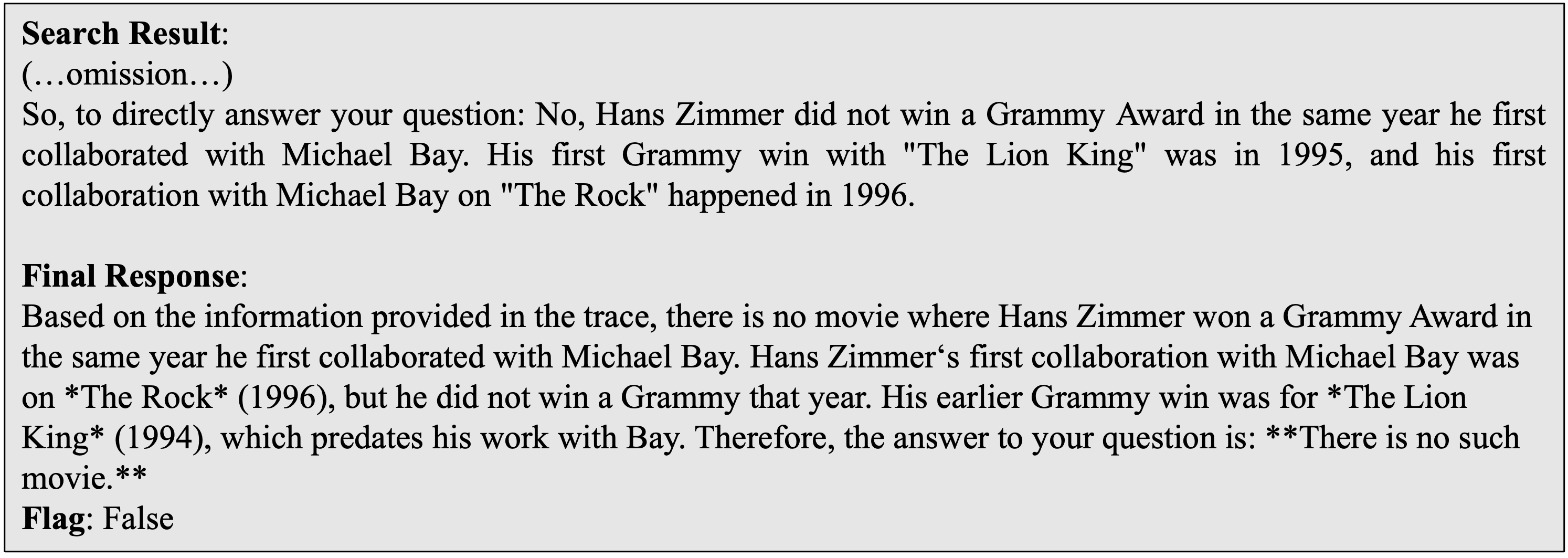} 
    \caption{The search results from Qwen Web Search}\label{fig:4}
\end{figure}

Qwen Web Search also attempted to analyze and summarize the search results, as shown in Figure \ref{fig:4}. However, it produced an incorrect result and did not present the original search results to the LLM. This omission significantly impeded the LLM's ability to derive the correct answer, resulting in an almost negligible chance of answering the question correctly.

In summary, the search results returned by Web Search Related MCP Servers and Web Search Related Function Calls directly influence whether LLM can accurately answer a given question. When search results are provided without any processing, LLMs must depend solely on their reasoning abilities to derive the final answer. On the other hand, when the results are analyzed and processed before being returned, LLMs can more easily obtain the answers. However, this approach places complete dependence on the server's ability to accurately respond to queries.

\subsection{Database Search Task}

For the Database Search task, we evaluate the search performance of PostgreSQL MCP Server and XiYan MCP Server using the SQL\_EVAL dataset. The example is shown in Figure \ref{fig:5}. Detailed examinations of additional cases are provided in Appendix \ref{app:database}.

\begin{figure}[h]
    \centering
    \includegraphics[width=1\textwidth]{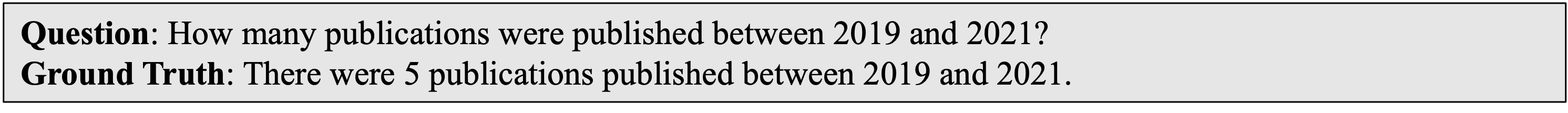} 
    \caption{The issue in the SQL\_EVAL dataset}\label{fig:5}
\end{figure}

\begin{figure}[h]
    \centering
    \includegraphics[width=1\textwidth]{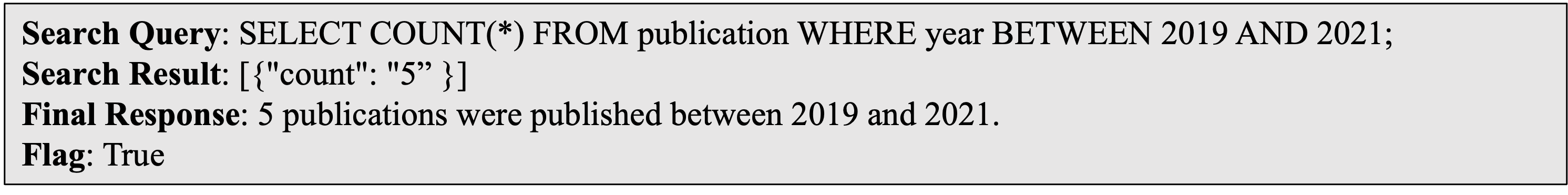} 
    \caption{The search results from PostgreSQL MCP Server}\label{fig:6}
\end{figure}

The PostgreSQL MCP Server operates by receiving an SQL query generated by the LLM in response to a specific question, as shown in Figure \ref{fig:6}. This SQL query is executed against the database, and the resulting data is returned as output. Subsequently, the LLM utilizes the database query results to formulate the final answer to the original question. In fact, the PostgreSQL MCP Server only handles database connections and executes SQL Query.

\begin{figure}[h]
    \centering
    \includegraphics[width=1\textwidth]{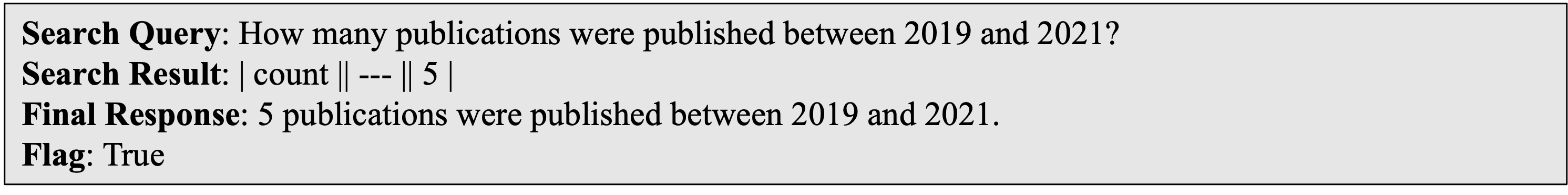} 
    \caption{The search results from XiYan MCP Server}\label{fig:7}
\end{figure}

Conversely, the XiYan MCP Server is designed to accept the original question directly as its input, as shown in Figure \ref{fig:7}. Within the server, the processes of SQL generation and execution are performed internally. As a result, the output from the XiYan MCP Server is the database query result, which the LM then uses to derive the final answer.

\section*{Conclusion}
The evaluation of various Model Context Protocol (MCP) servers highlights significant differences in both effectiveness and efficiency. While MCPs offer distinct advantages in structuring tool usage, they do not consistently demonstrate marked improvements over non-MCP approaches, such as function calls. Our experiments showed that the most effective MCP, Bing Web Search, achieved an accuracy of 64\%, whereas DuckDuckGo lagged at just 10\%. Furthermore, performance varied widely in terms of time consumption, with top performers like Bing and Brave Search executing tasks in under 15 seconds, contrasted with significantly slower alternatives like Exa Search.

Importantly, we found that the accuracy of MCP servers can be substantially enhanced by optimizing the parameters that LLMs must construct. For instance, transitioning from SQL-based queries to natural language processing in the XiYan MCP server resulted in a noteworthy increase in accuracy, demonstrating that incorporating a text-to-SQL model can lead to a 22 percentage point improvement.

Overall, while MCPs provide a structured means for AI tools to interact with data, there remains considerable potential for optimization. By addressing the challenges LLMs encounter in parameter construction and enhancing the user-friendliness of tool interfaces, developers can significantly improve the performance and reliability of MCP servers. This research paves the way for further investigations into optimized MCP implementations, ultimately leading to better AI-driven search and data retrieval solutions.

\bibliographystyle{plain}  
\bibliography{ref} 

\newpage

\appendix\

\section{Dataset}
In this section, we will introduce the construction methods and provide example demonstrations of the non-public datasets, including News (Chinese) and Knowledge (Chinese) for the Web Search Task, and the Car\_bi dataset for the Database Search Task.

\begin{figure}[h]
    \centering
    \includegraphics[width=1\textwidth]{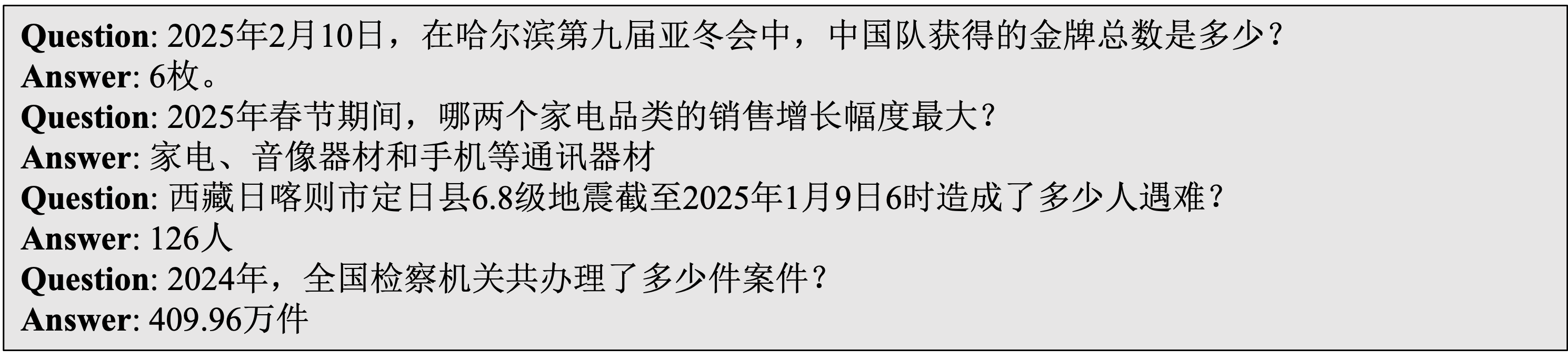} 
    \caption{Samples of the News (Chinese) Dataset}\label{fig:19}
\end{figure}

\textbf{News (Chinese) Dataset} is collected by cleaning data from daily Xinwen Lianbo transcripts over the past three months and processing it using reverse engineering techniques. The code will be open-sourced later. Dataset examples see Figure \ref{fig:19}.

\begin{figure}[h]
    \centering
    \includegraphics[width=1\textwidth]{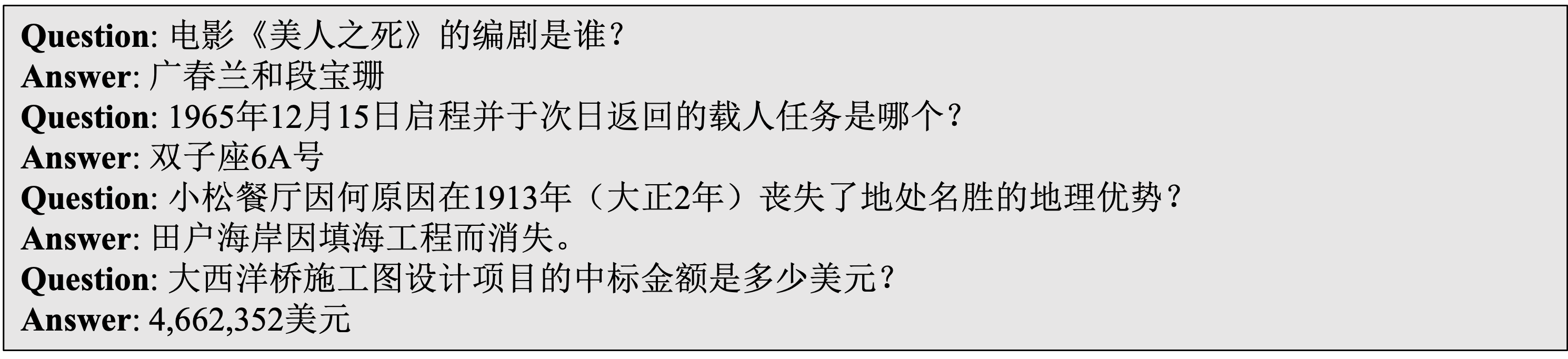} 
    \caption{Samples of the Knowledge (Chinese) Dataset}\label{fig:20}
\end{figure}

\textbf{Knowledge (Chinese) Dataset} is collected by cleaning data from knowledge-intensive websites like Wikipedia and science and technology reports, and processing it using reverse engineering techniques. The code will be open-sourced later. Dataset examples see Figure \ref{fig:20}.

\begin{figure}[!h]
    \centering
    \includegraphics[width=1\textwidth]{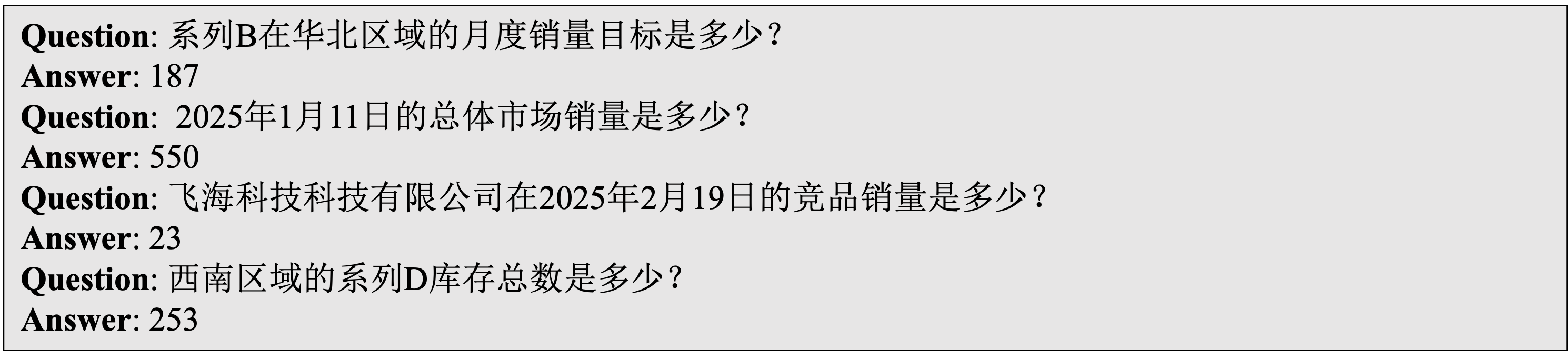} 
    \caption{Samples of the Knowledge (Chinese) Dataset}\label{fig:21}
\end{figure}

\textbf{Car\_bi (Chinese) Dataset} is a synthetic dataset from an automobile manufacturer data source. Dataset examples see Figure \ref{fig:21}.

\begin{figure}[!h]
    \centering
    \includegraphics[width=1\textwidth]{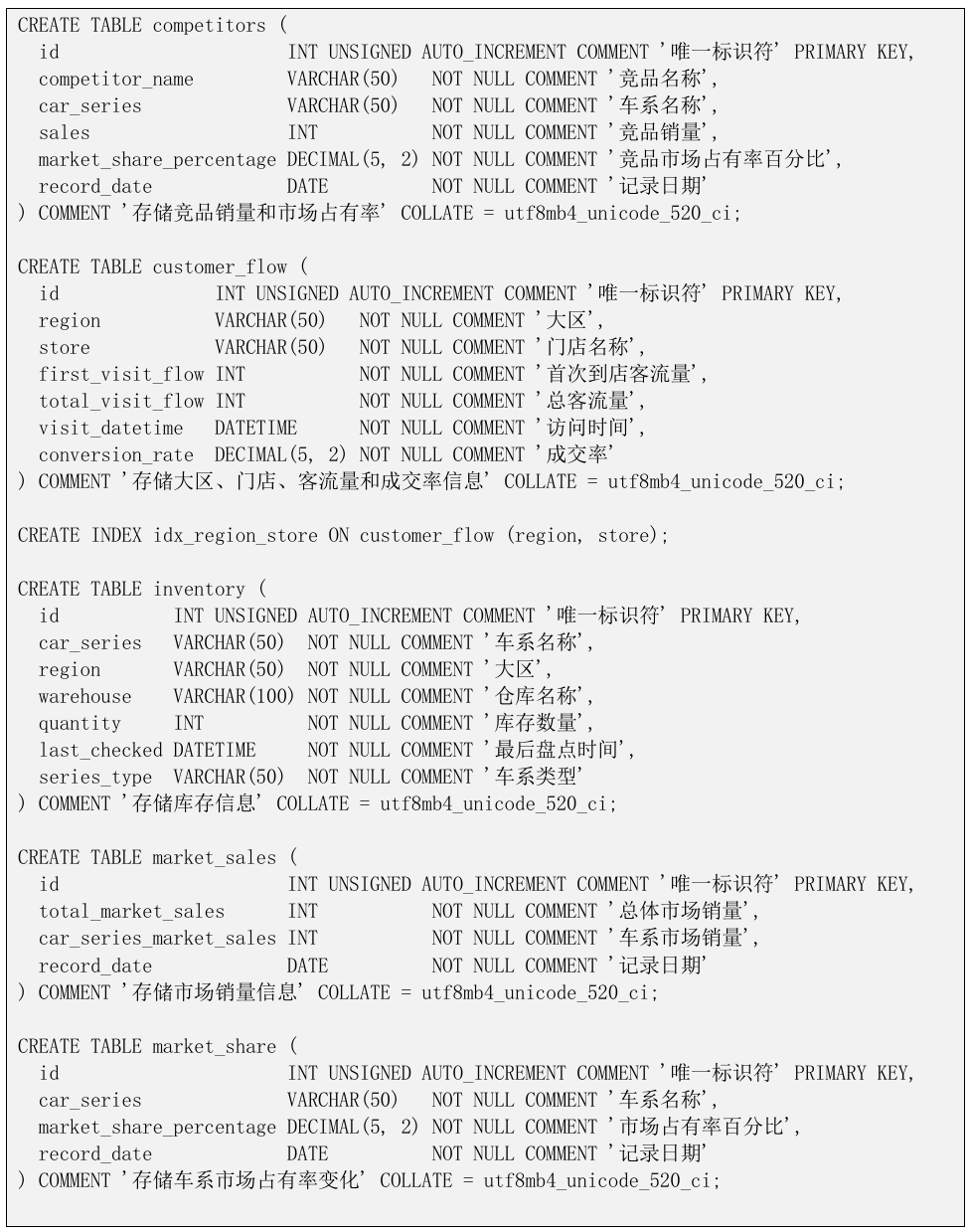} 
    \caption{The DDL of the Car\_bi Dataset}\label{fig:22}
\end{figure}

The DDL of the Car\_bi Dataset is shown in the Figure \ref{fig:22}. Due to space limitations, only a portion of the DDL information is presented.

\section{Web Search Task}\label{app:web}
\begin{figure}[h]
    \centering
    \includegraphics[width=1\textwidth]{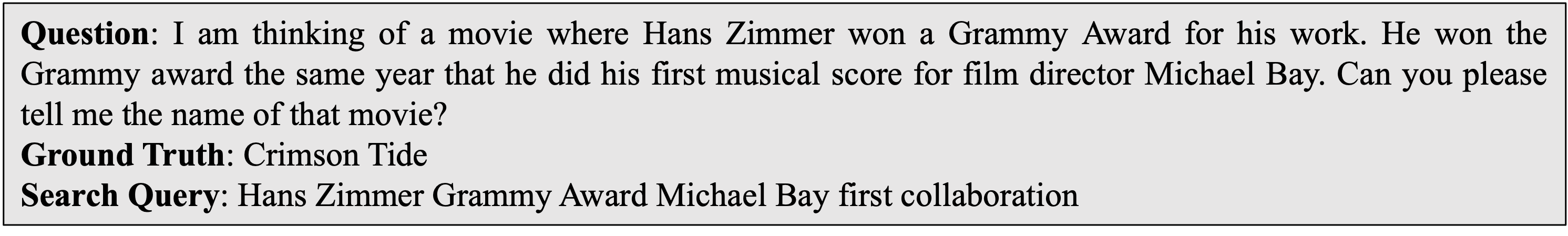} 
    \caption{The issue in the Frames dataset}\label{fig:88}
\end{figure}

In this section, we conduct a comparative analysis of search results within the Web Search task using the Frames dataset, as shown in Figure \ref{fig:88}. Specifically, we examine the performance of various MCP Servers and Function Calls, excluding Brave Search, BochaAI, and Qwen Web Search. 

\subsection{Web Search Related MCP}

\begin{figure}[h]
    \centering
    \includegraphics[width=1\textwidth]{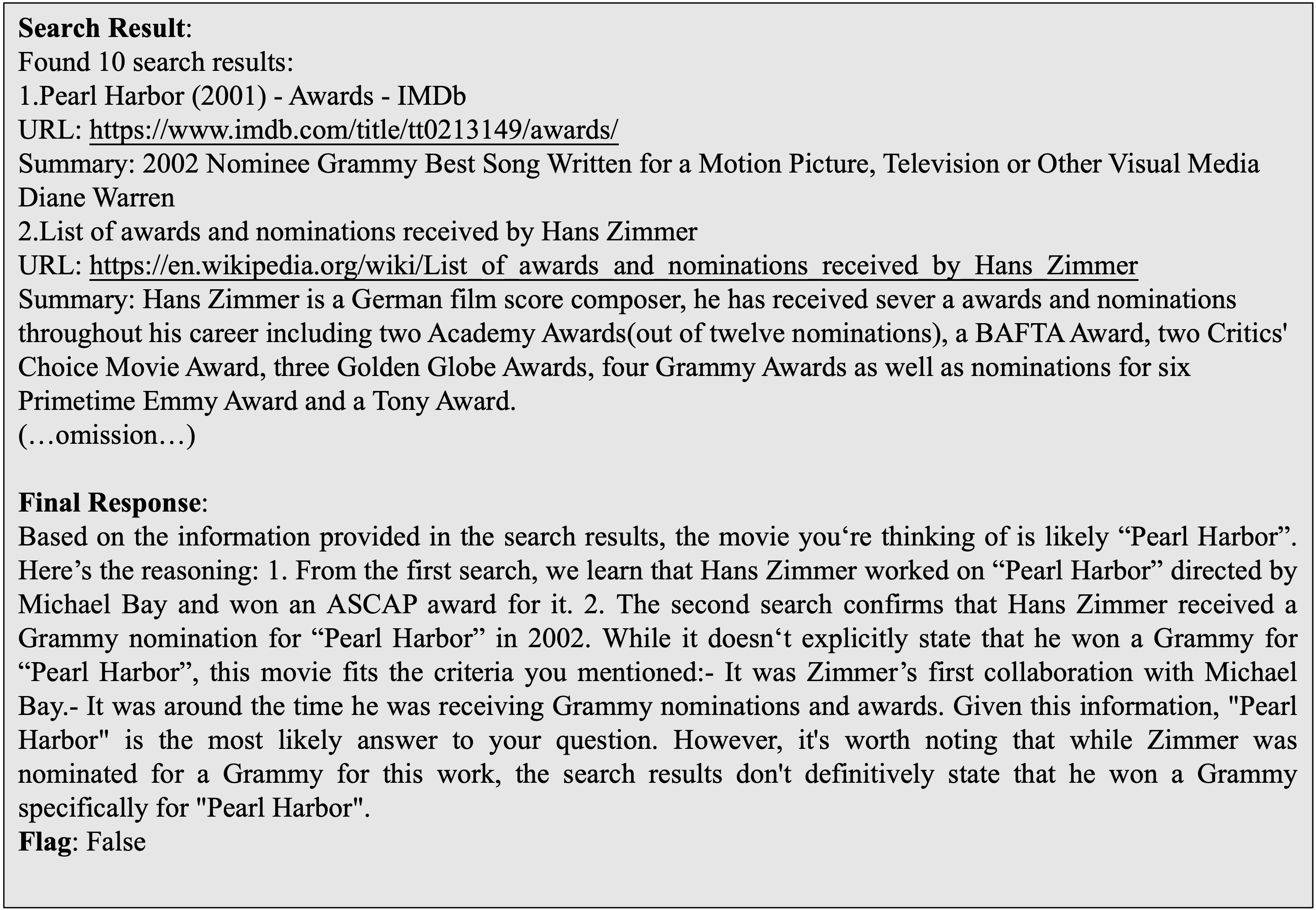} 
    \caption{The search results from DuckDuckGo Search Server}\label{fig:9}
\end{figure}

The DuckDuckGo Search Server returns the top ten web query results related to the question, including each webpage's title, URL, and summary, as shown in Figure \ref{fig:9}. The results include some sub-questions related to the main question but do not contain all the information needed to solve the issue. There may even be misleading information, such as the top result being "Pearl Harbor," making it difficult for the LLM to deduce the correct answer from so much information.

\begin{figure}[h]
    \centering
    \includegraphics[width=1\textwidth]{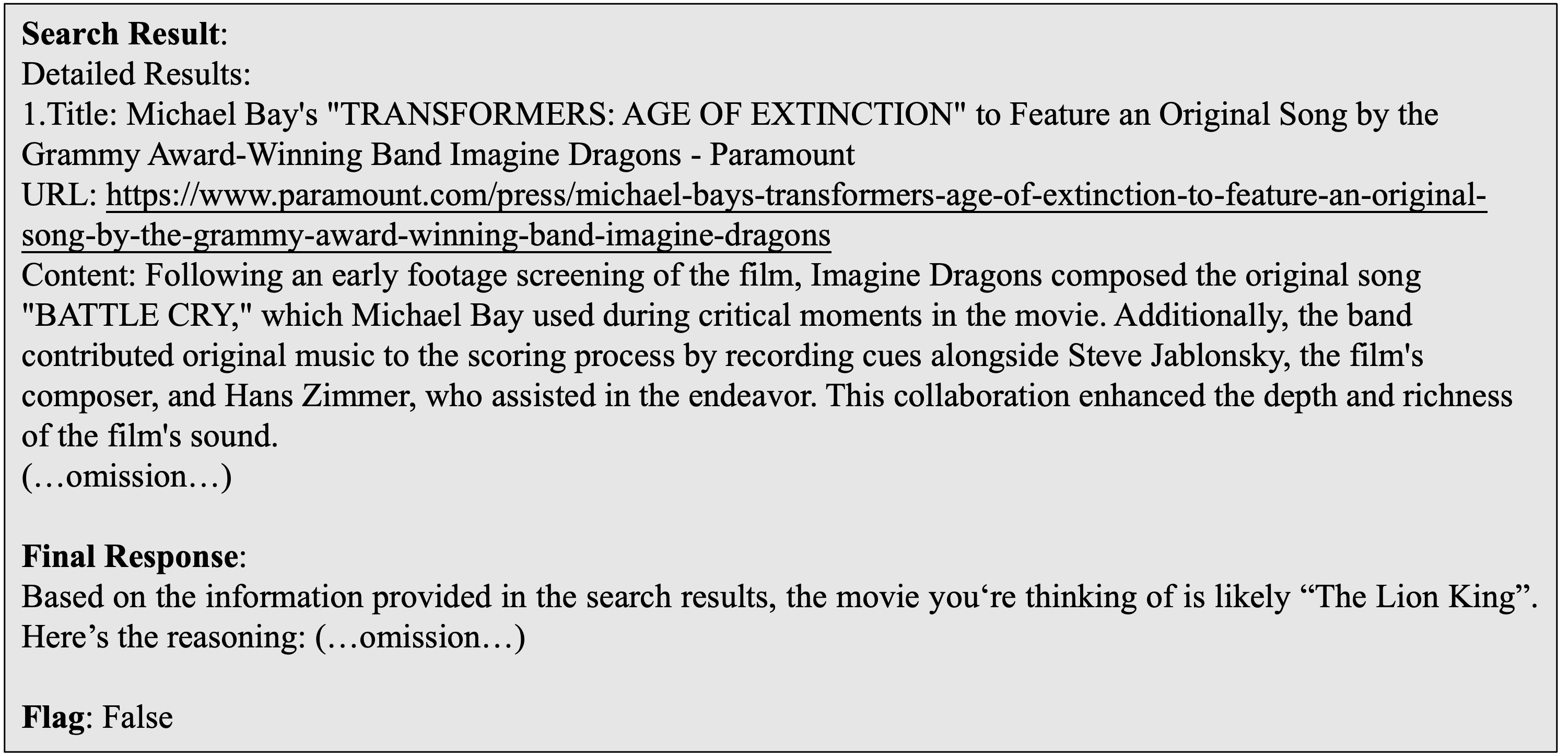} 
    \caption{The search results from Tavily MCP Server}\label{fig:10}
\end{figure}

The Tavily MCP Server returns the top five web query results related to the question, including each webpage's title, URL, and content, as shown in Figure \ref{fig:10}. Instead of providing the original webpage content, the Tavily MCP Server formats and summarizes the content. However, because the returned webpage content is not directly related to the original question, the LLM is unable to obtain the correct answer.

\begin{figure}[h]
    \centering
    \includegraphics[width=1\textwidth]{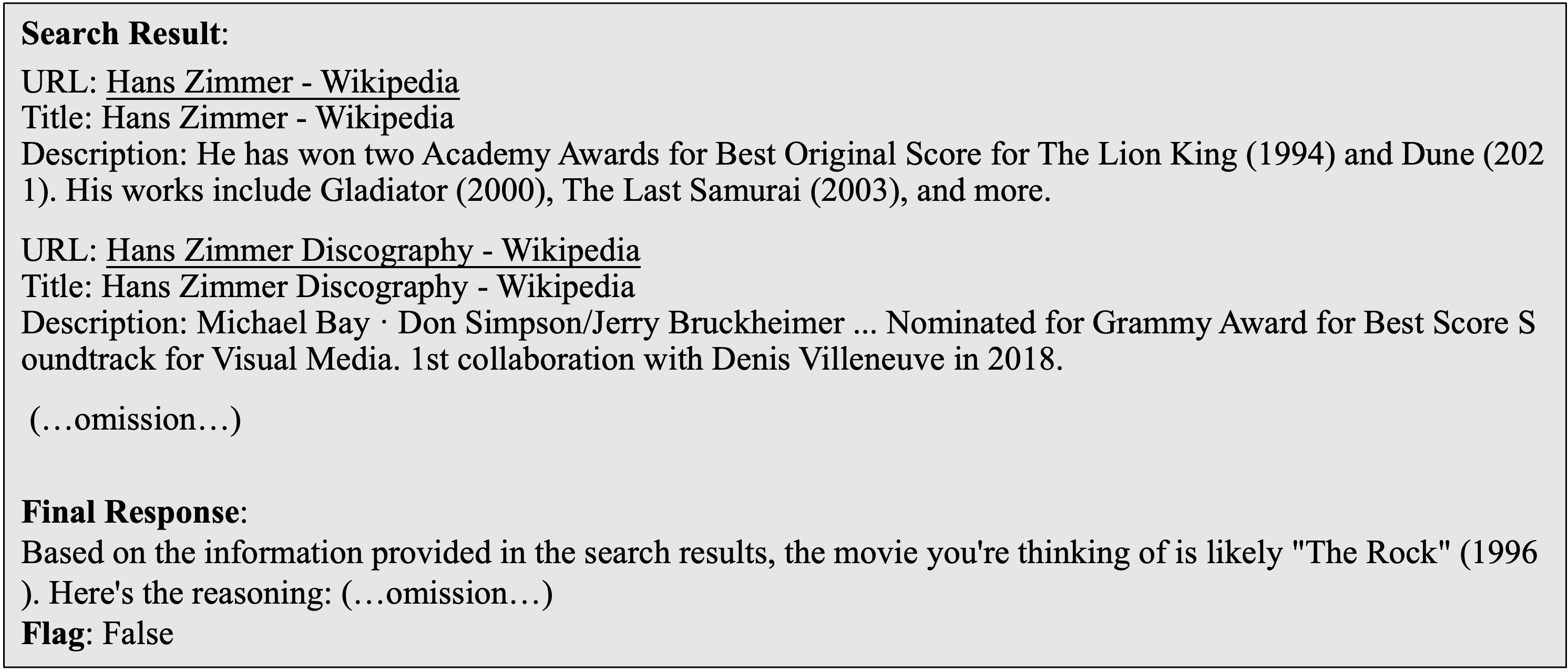} 
    \caption{The search results from Fire Crawl Search}\label{fig:12}
\end{figure}

Fire Crawl Search returned the top five Wikipedia URLs, titles, and descriptions related to the question, as shown in Figure \ref{fig:12}. Each description contains only the main information related to the title, but this information is completely insufficient for the LLM to answer the question. More information may be included in the main text, but Fire Crawl Search did not return it, so it is necessary to visit the URLs again to obtain the complete information.

\begin{figure}[h]
    \centering
    \includegraphics[width=1\textwidth]{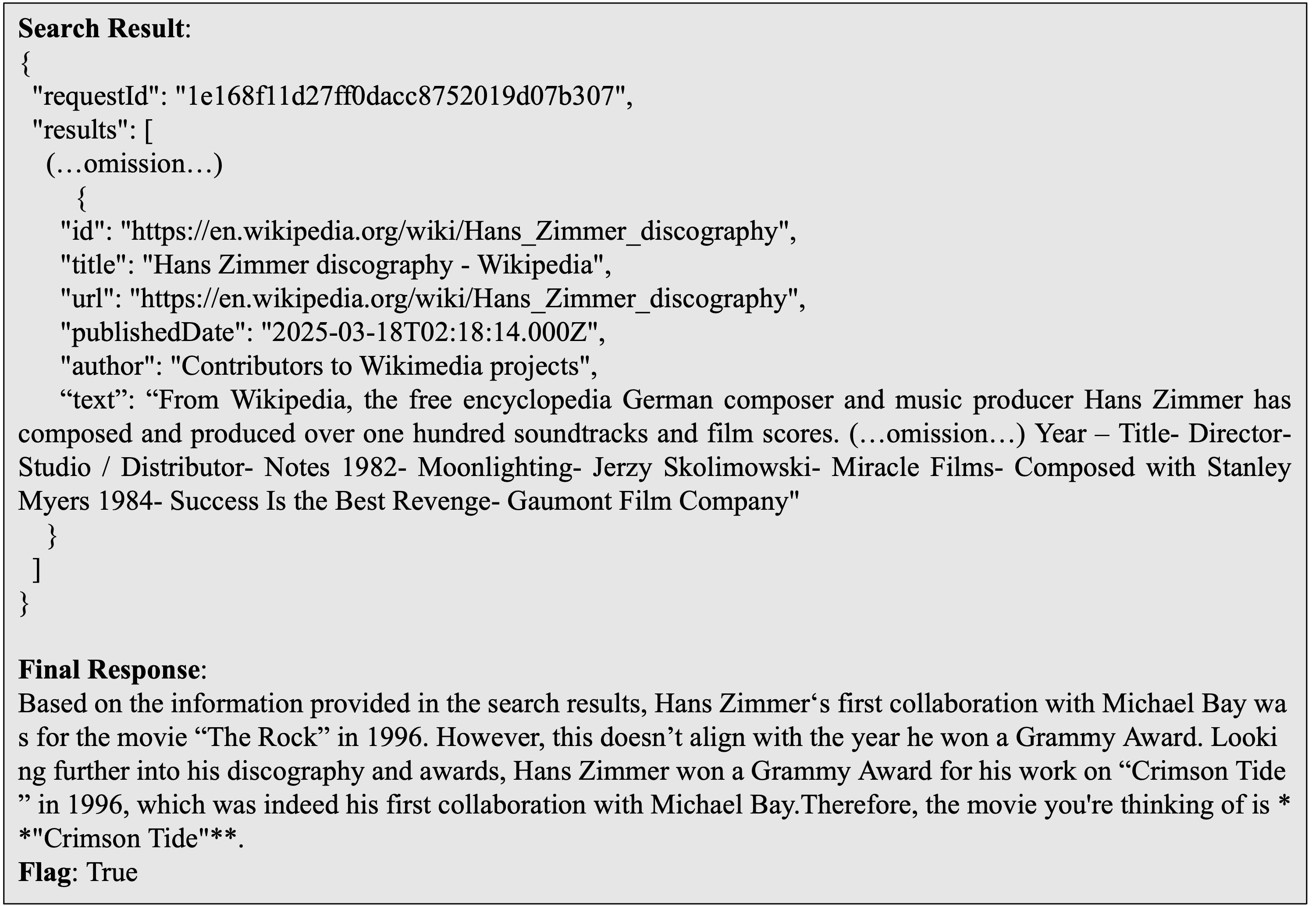} 
    \caption{The search results from Exa Search}\label{fig:11}
\end{figure}

Exa Search returns a JSON-formatted result that includes a requestId and results, as shown in Figure \ref{fig:11}. The results are a list of dictionaries, each containing the id, title, url, publishedDate, author, and text of the returned item. The text provides a summary and analysis of the page. Therefore, using this result, the LLM can easily obtain the correct answer without needing to parse and analyze complex webpage structures.

\begin{figure}[!h]
    \centering
    \includegraphics[width=1\textwidth]{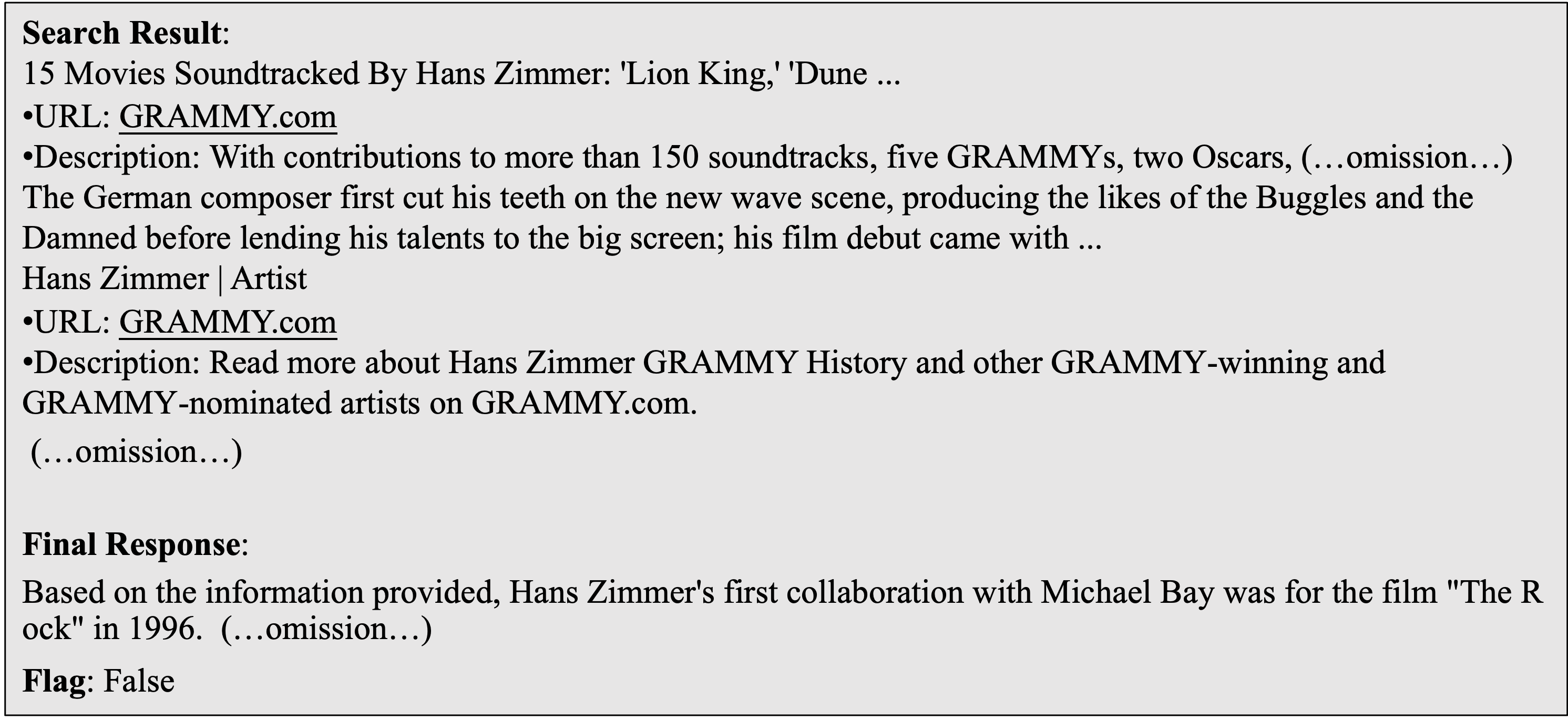} 
    \caption{The search results from Bing Web Search}\label{fig:13}
\end{figure}

Bing Web Search returns query results for sub-questions related to the main question, including each sub-question's description (e.g. Hans Zimmer| Artist), URL, and description, as shown in Figure \ref{fig:13}. Although they originate from the same website, they address different questions. The description provides an answer to the sub-question. However, it is important to note that Bing Web Search automatically truncates information beyond a certain point in the returned results, which may contain the correct answer.

\newpage
\subsection{Web Search Related Function Calls}
\begin{figure}[h]
    \centering
    \includegraphics[width=1\textwidth]{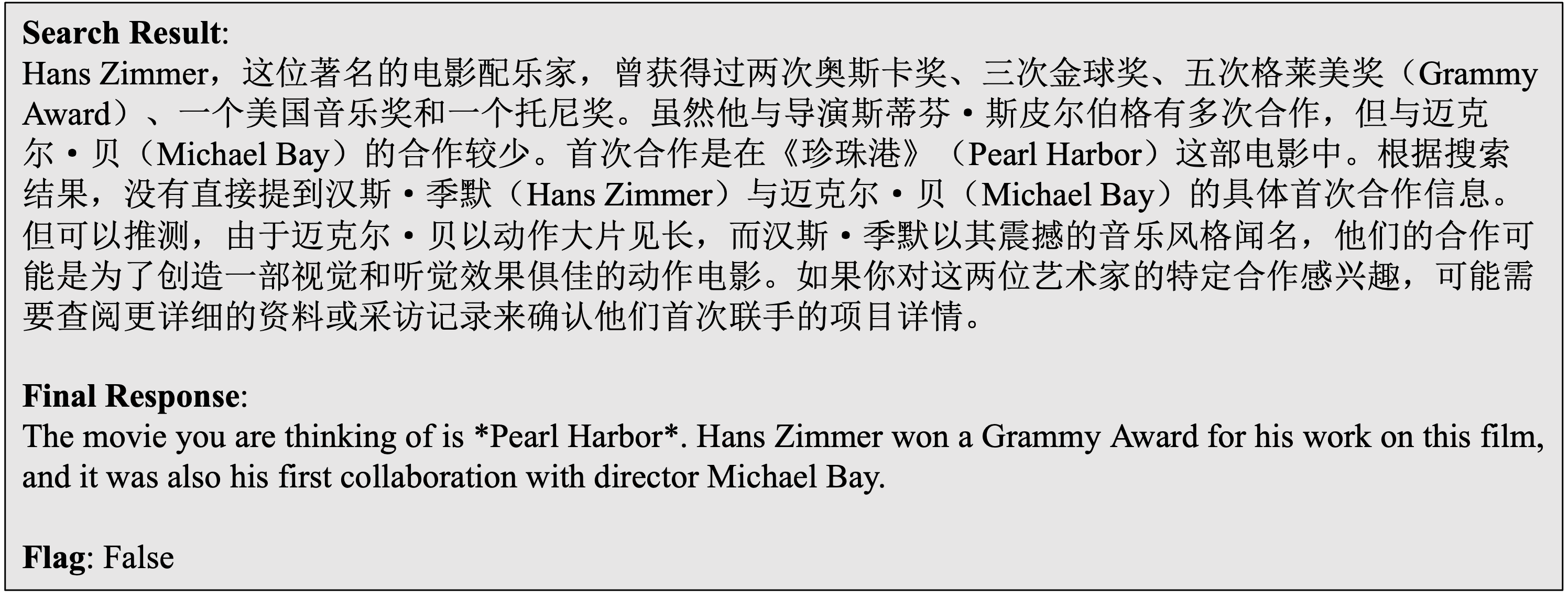} 
    \caption{The search results from Quark Search}\label{fig:14}
\end{figure}

Quark Search directly returns an answer to the question, as shown in Figure \ref{fig:14}. Quark Search itself handles the analysis of the search results and responds to the query, allowing the LLM to use the returned results without needing to perform its own reasoning. However, when Quark Search makes an error in its analysis, the LLM cannot access the original search results, making it almost impossible to obtain the correct outcome.

\section{Database Search Task}\label{app:database}
\begin{figure}[h]
    \centering
    \includegraphics[width=1\textwidth]{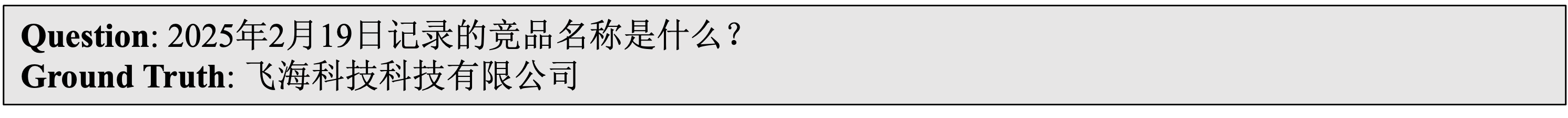} 
    \caption{The issue in the car\_bi dataset}\label{fig:15}
\end{figure}

In this section, we compare the performance of the MySQL MCP Server and the XiYan MCP Server on the car\_bi dataset. The example is shown in Figure \ref{fig:15}.

\begin{figure}[h]
    \centering
    \includegraphics[width=1\textwidth]{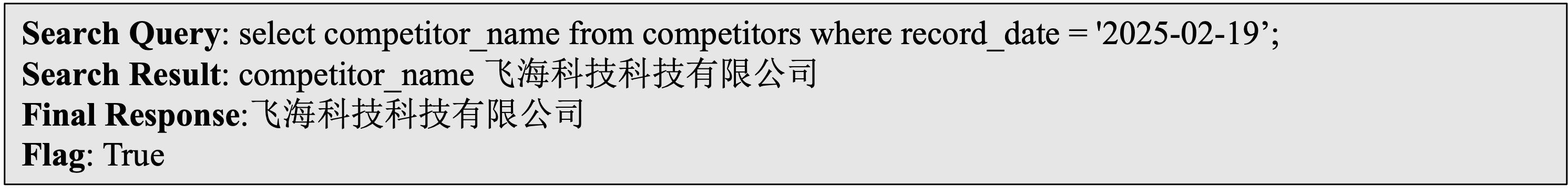} 
    \caption{The search results from MySQL MCP Server}\label{fig:16}
\end{figure}

The purpose of the MySQL MCP Server is to connect to the database and execute database queries, as shown in Figure \ref{fig:16}. For this issue, the LLM needs to generate an SQL query that addresses the problem and provide it as input to the MySQL MCP Server. The server executes the SQL query and returns the results, and the LLM generates the final answer based on those query results.

\begin{figure}[!h]
    \centering
    \includegraphics[width=1\textwidth]{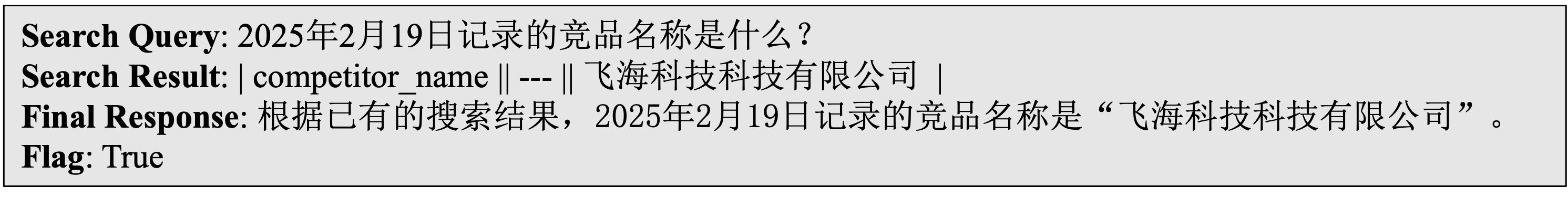} 
    \caption{The search results from XiYan MCP Server}\label{fig:17}
\end{figure}

The input to the XiYan MCP Server is the original question, and the output is the final answer, as shown in Figure \ref{fig:17}. XiYan MCP Server can generate SQL queries from original questions and execute database queries to obtain the final results, and then return to the LLM. 

\end{document}